\documentclass{iau}
\usepackage{graphicx}
\usepackage{amssymb}

\usepackage{braket}

\def\zs{\mbox{{$z_{\rm spec}$}}}
\def\zp{\mbox{{$z_{\rm phot}$}}}

\title[METAPHOR: PDF's for machine learning photo-z]{METAPHOR:  Probability density estimation for machine learning based photometric redshifts}
\author[Amaro et al. 2017]{V.~Amaro$^{1}$, S.~Cavuoti$^{2}$, M.~Brescia$^{2}$, C.~Vellucci$^{3}$, C.~Tortora$^{4}$,  G.~Longo$^{1}$}
\affiliation{
$^{1}$Dept. of Physical Sciences, University of Napoli Federico II, via Cinthia 9, 80126 Napoli, Italy\\[\affilskip]
$^{2}$INAF - Astronomical Observatory of Capodimonte, via Moiariello 16, 80131 Napoli, Italy\\[\affilskip]
$^{3}$DIETI, University of Naples Federico II, Via Claudio,21 I-80125 Napoli, Italy\\[\affilskip]
$^{4}$Kapteyn Astronomical Institute, Univ. of Groningen,  9700 AV Groningen, the Netherlands}

\pubyear{2017}
\volume{325}  %% insert here IAU Symposium No.
\jname{Astroinformatics 2016}
\editors{M. Brescia, S.G. Djorgovski, E. Feigelson, \\  G. Longo \& S. Cavuoti, eds.}

\begin{document}

\maketitle

\begin{abstract}

We present METAPHOR (Machine-learning Estimation Tool for Accurate PHOtometric Redshifts), a method able to provide a reliable PDF for photometric galaxy redshifts estimated through empirical techniques.
METAPHOR is a modular workflow, mainly based on the MLPQNA neural network as internal engine to derive photometric galaxy redshifts, but giving the possibility to easily replace MLPQNA with any other method to predict photo-z's and their PDF.
We present here the results about a validation test of the workflow on the galaxies from SDSS-DR9, showing also the universality of the method by replacing MLPQNA with KNN and Random Forest models.
The validation test include also a comparison with the PDF's derived from a traditional SED template fitting method (Le Phare).

\end{abstract}

\firstsection % if your document starts with a section,
\section{Introduction}
Galaxy redshifts are important for a large number of studies related with the extragalactic universe, due to their direct correlation with the distance of the sources. Photometric galaxy redshifts (hereafter photo-z's) are crucial in the current era of large surveys, based on massive datasets. They are used in a wide plethora of tasks, such as, for example, to constrain the dark matter and dark energy contents of the Universe through weak gravitational lensing, to understand the cosmic large scale structure, by identifying galaxies clusters and groups, to map the galaxy color-redshift relationships, as well as to classify astronomical sources.
More recently, the attention in this field has been focused on the techniques able to compute a Probability Density Function (PDF) of the photo-z's for each individual astronomical source, with the goal to improve the knowledge about statistical reliability of photo-z estimations.
In the machine learning context several methods have been proposed to approach this task, see for instance: (\cite[Bonnet 2013]{bonnet2013}, \cite[Rau et al. 2015]{rau2015}, \cite[Sadeh et al. 2015]{sadeh2015}, \cite[Carrasco \& Brunner 2014]{carrasco2014b}).
Here we present a new method, named METAPHOR (Machine-learning Estimation Tool for Accurate PHOtometric Redshifts), a modular workflow including a machine learning engine to derive photo-z's and a method to produce their PDF's, based on the evaluation of photometric data uncertainties to derive a perturbation law of the photometry.
With this law we perform the perturbation of the features, in a controlled, not biased by systematics, way. A proper error fitting, accounting for the attribute errors, allows to constrain the perturbation of photometry on the biases of the measurements.

\section{The METAPHOR workflow}%\label{SEC:metaphor}

The conceptual flow of the METAPHOR pipeline is based on the following sequence of tasks: given a multi-band data sample containing the spectroscopic galaxy redshifts, \textit{(i)} for each band involved, a photometry perturbation function is derived; \textit{(ii)} the data sample is randomly shuffled and split into a training and a test set; \textit{(iii)} the photometry of the test set is perturbed, thus obtaining an arbitrary number $N$ of test set replica; \textit{(iv)} finally, the machine learning engine is trained and the $N+1$ test sets ($N$ perturbed plus the unperturbed one) are submitted to the training model to derive the PDF of photo-z estimations.

In the last step, the $N+1$ values, output of the trained network, are used to calculate, for each bin of redshift, the probability that a given photo-z value belongs to each bin. The binning step $B$, as well as the number $N$ of perturbations, are user defined parameters, to be chosen accordingly to the specific requirements of the experiment. For a given photo-z binning step $B$, we calculate the number of photo-z's for each bin ($C_{B,i}\  \in [Z_{i}, Z_{i+B}[$) and the probability that the redshift belongs to the bin is $P(Z_{i} \leq \mbox{Photo-z} < Z_{i+B}) = C_{B,i}/(N+1)$. The resulting PDF is thus formed by all these probabilities. 

At the end of the procedure, a post-processing module calculates the final photo-z estimation and PDF statistics. For instance, we evaluate the photo-z's in terms of a standard set of statistical estimators for the quantity $\Delta z = (\zs-\zp)/(1+\zs)$ on the objects in the blind test set: \textit{(a)} bias: defined as the mean value of the residuals $\Delta z$; \textit{(b)} $\sigma$: the standard deviation of the residuals; \textit{(c)} $\sigma_{68}$: the radius of the region that includes $68\%$ of the residuals close to 0;
\textit{(d)} $NMAD$: the Normalized Median Absolute Deviation of the residuals, defined as $NMAD(\Delta z) = 1.48 \times Median (|\Delta z|)$; \textit{(e)} fraction of outliers with $|\Delta z| > 0.15$; \textit{(f)} skewness: asymmetry of the probability distribution of a real-valued random variable around the mean.

Furthermore, in order to evaluate the cumulative performance of the PDF we compute the following three estimators on the \textit{stacked} residuals of the PDF's: \textit{(1)} $f_{0.05}$: the percentage of residuals within $\pm 0.05$; \textit{(2)} $f_{0.15}$: the percentage of residuals within $\pm 0.15$; \textit{(3)} $\Braket{\Delta z}$: the weighted average of all the residuals of the \textit{stacked} PDF's.

The photometry perturbation is based on the following expression, applied on the given \textit{j} magnitudes of each band \textit{i} as many times as the number of perturbations of the test set:
\begin{equation}
 m_{ij}=m_{ij}+\alpha_{i}K_{ij}*gaussRandom_{(\mu=0,\sigma=1)}
\end{equation}
The term $\alpha_i$ is a multiplicative constant, used to customize the photometric error trend on the base of the specific band photometric quality. This could result particularly useful in case of photometry obtained by merging different surveys; the quantity $K_{ij}(x)$ is the weighting coefficient associated to each specific band used to weight the Gaussian noise contribution to magnitude values; finally, the term $gaussRandom_{(\mu=0,\sigma=1)}$ is a random value extracted from a normal distribution.

We investigated four different types of the weighting coefficient $K_{ij}(x)$. First one is a heuristically chosen real number between $0$ and $1$, implying a same width of the gaussian noise for each point. The second choice is based on weighting the Gaussian noise contribution using the individual magnitude error provided for each source. The third one is a polynomial fitting: a binning of photometric bands is performed, in which a polynomial fitting of the mean magnitude errors is used to reproduce the intrinsic trend of the distribution. The last option is a slightly more sophisticated version of the polynomial fitting, coupled with a minimum value chosen heuristically, thus resulting in a bi-modal perturbation function.

\section{The experiments}

As introduced, one of the most suitable features of METAPHOR is the invariance to the specific empirical model used as engine to estimate photo-z's.  In order to demonstrate this capability, we tested the METAPHOR workflow using three different machine learning methods: MLPQNA neural network \cite[(Byrd et al. 1994)]{byrd1994}, already successfully used in several astrophysical contexts (\cite[Brescia et al. 2013]{brescia2013}, \cite[Brescia et al. 2014a]{dame}, \cite[Cavuoti et al. 2012]{cavuoti2012}, \cite[Cavuoti et al. 2014a]{cavuoti2014}, \cite[Cavuoti et al. 2014b]{cavuoti2014b}, \cite[Cavuoti et al. 2015b]{cavuoti2015}), the standard KNN \cite[(Cover \& Hart 1967)]{cover1967}, and Random Forest \cite[(Breiman 2001)]{breiman2001}. In particular, the experiment with a very basic machine learning model like KNN method, would demonstrate the most general applicability of any empirical model engine within METAPHOR.
Furthermore, by considering that the methods mostly based on SED template fitting intrinsically provide the PDF of the estimated photo-z's, we compared METAPHOR with the \textit{Le Phare} model \cite[(Ilbert et al. 2016)]{ilbert2006}.

The real data used for the tests were a galaxy spectroscopic catalogue sample extracted from the Data Release 9 (DR9) of the Sloan Digital Sky Survey, SDSS, \cite[(York et al 2000)]{sdss}. By using MLPQNA as internal engine for photo-z estimation, we reached values of $\sigma = 0.024$, $bias \sim 0.0063$ and $\sim 0.12\%$ of outliers. These statistical results are slightly worse than what we showed in a previous article \cite[(Brescia et al. 2014c)]{brescia2014c}, where we already used the MLPQNA method to derive photo-z's for the galaxies of SDSS-DR9. However, this discrepancy is only apparent, by considering that the spectroscopic KB used in the cited work was much larger than the one used here ($\sim 150,000$ training objects against only $\sim 50,000$ objects used for the training in this case). The decision of such limited sample used in the present experiment was induced by the different goal of the experiment.

We performed a large number of experiments with MLPQNA using $100$ photometric perturbations in order to find the best perturbation law. The most performing experiment turns out to be the one based on a bi-modal perturbation law with threshold $0.05$ and a multiplicative constant $\alpha=0.9$. This experiment leads to a stacked PDF with $\sim92\%$  within $[-0.05, 0.05]$, $\sigma_{68}=0.019$, $\sim21\%$ of the objects falling within the peak of the PDF, $\sim53\%$ falling within $1$ bin from the peak and $\sim82\%$ falling within the PDF. 

After having found the best perturbation law, we executed $1000$ perturbations of the test set. This experiment led to an increase in the statistical performances, obtaining $\sigma_{68}=0.018$ and $\sim21.8\%$ within the peak of the PDF, $\sim54.4\%$ within $1$ bin from the peak and $\sim89.6\%$ inside the PDF.

The same configuration and perturbed data have been used to estimate photo-z's by replacing MLPQNA with, respectively, the KNN and Random Forest models within the METAPHOR workflow. In parallel, we derived also the photo-z PDF's with the \textit{Le Phare} method. The statistical results for all these methods are summarized in Table~\ref{tab:stackedstat}.

\begin{table}
 \centering
 \begin{tabular}{ccccc}
Estimator	    	& MLPQNA 	& KNN       & RF        & \textit{Le Phare}     	\\ \hline
$bias$		    	& $0.0006$	& $0.0029$  & $0.0035$  &  $0.0009$	\\
$\sigma$	    	& $0.024$	& $0.026$   & $0.025$   &  $0.060$	\\
$\sigma_{68}$   	& $0.018$	& $0.020$   & $0.019$   &  $0.035$     	\\
$NMAD$ 		   	    & $0.017$	& $0.018$   & $0.018$   &  $0.030$     	\\
$skewness$	   	    & $-0.17$	& $0.330$   & $0.015$   &  $-18.076$   	\\
$outliers>0.15$	    & $0.11\%$	& $0.15\%$  & $0.15\%$  &  $0.69\%$    \\ \hline
 $f_{0.05}$		        & $91.7\%$	& $92.0\%$   & $92.1\%$   & $71.2\%$   \\
 $f_{0.15}$		        & $99.8\%$	& $99.8\%$   & $99.7\%$   & $99.1\%$   \\
 $\Braket{\Delta z}$	& $-0.0006$	& $-0.0018$  & $-0.0016$  & $0.0131$   \\ \hline
 \end{tabular}
\caption{Statistical results of photo-z's and related PDF estimation on the blind test set extracted from SDSS-DR9, obtained by three machine learning models (MLPQNA, KNN and Random Forest), alternately used as internal engine of METAPHOR and by the SED template fitting method \textit{Le Phare}. The last three estimators are related to the cumulative PDF of the estimated photo-z's. See text for the explanation of the statistical estimators.} \label{tab:stackedstat}
\end{table}

Although there is a great difference in terms of photo-z estimation statistics between \textit{Le Phare} and MLPQNA (see Table~\ref{tab:stackedstat}), the results of the PDF in terms of $f_{0.15}$ are comparable. But the greater efficiency of MLPQNA induces an improvement in the range within $f_{0.05}$, where we find $\sim 92\%$ of the objects against the $\sim 72\%$ for \textit{Le Phare}. Both individual and \textit{stacked} PDF's are more symmetric in the case of empirical methods. This is particularly evident by observing the skewness (see Table \ref{tab:stackedstat}), which is $\sim 100$ times greater in the case of \textit{Le Phare}.

The presented photo-z estimation results and the statistical performance of the cumulative PDF's, achieved by MLPQNA, RF and KNN through the proposed workflow, demonstrate the validity and reliability of the METAPHOR strategy, despite its simplicity, as well as its general applicability to any other empirical method.

\section*{Acknowledgments}
MB and SC acknowledge financial contribution from the agreement ASI/INAF I/023/12/1.
MB acknowledges the PRIN-INAF 2014 \textit{Glittering kaleidoscopes in the sky: the
multifaceted nature and role of Galaxy Clusters}.
CT is supported through an NWO-VICI grant (project number $639.043.308$).

\end{document}